# Hydrocarbons Heterogeneous Pyrolysis: Experiments and Modeling for Scramjet Thermal Management

Marc BOUCHEZ*, Emeric DANIAU†,
*MBDA-F, 8 rue Le Brix, 18020 Bourges CEDEX – France*

Nicolas VISEZ, Olivier HERBINET‡, René FOURNET¦ and Paul Marie MARQUAIRE§,
*DCPR, UNR 7630 CNRS-INPL, 1 rue Grandville, 54000 Nancy – France.*

**The last years saw a renewal of interest for hypersonic research in general and regenerative cooling specifically, with a large increase of the number of dedicated facilities and technical studies. In order to quantify the heat transfer in the cooled structures and the composition of the cracked fuel entering the combustor, an accurate model of the thermal decomposition of the fuel is required. This model should be able to predict the fuel chemical composition and physical properties for a broad range of pressures, temperatures and cooling geometries. For this purpose, an experimental and modeling study of the thermal decomposition of generic molecules (long-chain or polycyclic alkanes) that could be good surrogates of real fuels, has been started at the DCPR laboratory located in Nancy (France). This successful effort leads to several versions of a complete kinetic model. These models do not assume any effect from the material that constitutes the cooling channel. A specific experimental study was performed with two different types of steel (regular: E37, stainless: 316L). Some results are given in the present paper.**

* Aerospace engineer, AIAA member,
† Aerospace engineer, AIAA member,
‡ Ph.D,
¦ University Professor,
§ Senior CNRS researcher

# I. Introduction

One of the main issues of the development of scramjet, an air breathing engine that could be used for hypersonic flight, is the thermal management of the vehicle and more especially of the engine[1,2]. Studies are currently conducted in France by ONERA and MBDA-France in order to develop an active cooling system that consists in using the fuel heat sink capability due to the endothermicity of its thermal decomposition. Because of the large heat load found in a scramjet, engine temperature is expected to increase well above 800 K leading in the thermal decomposition of the fuel circulating at the engine walls before the injection in the combustion chamber. A consequence of the decomposition of the heavy hydrocarbon fuel is the production of lighter species and of poly-aromatic compounds. In order to quantify the heat transfer through the walls of the engine and the composition of the fuel entering the combustion chamber, a detailed kinetic model of the thermal decomposition of the fuel is required.

To answer this problem, ONERA developed a plug reactor using a relatively large fuel flow rate[3], and MBDA-F collaborated with the DCPR laboratory to study the kinetics of the decomposition of fuels in a small perfectly-stirred reactor (PSR).

The thermal decomposition of n-dodecane has been studied in a jet-stirred reactor over a wide range of temperatures from 793 K to 1073 K and for residence times between 1 and 5 s. 32 products of the reaction have been analyzed. Major products are ethylene, methane and hydrogen (low mass fraction but high mole fraction). Other products are mainly 1-alkenes from propene to 1-undecene and aromatic compounds. The formation of aromatic compounds intervenes rather quickly; it takes importance for conversions higher than 25 %.

A detailed kinetic model generated by the EXGAS software has been completed by a set of elementary reactions involving aromatic and poly-aromatic compounds. The new model obtained is made of 1449 reactions and contains 271 species[4]. Agreement between experimental and computed results is satisfactory especially for primary products such 1-alkenes. As far as aromatic compounds are concerned a light variation between experiments and modeling is observed for long residence times especially. This may be related to uncertainties on kinetic parameters of reactions involved in the formation of benzene ($C_5$-pathway) and poly-aromatics compounds. To expand our knowledge on the pyrolysis of other hydrocarbons, some cyclanes (cyclopentane, cyclohexane and norbornane) were also studied in our jet-stirred reactor[5].

It was found that the thermal stability of cyclanes is generally better than its linear alkane counterpart and that some differences could be seen in the decomposition products distribution.

Light species (hydrogen, methane, ethylene, ethane, propene) fractions are very similar while middleweight 1-alkenes (from 1-butene to 1-undecene) account for a large part of the n-dodecane decomposition products (at small conversion rate), 1,3-butadiene plays the same role for cyclohexane.

At higher conversion rate, ethylene becomes the major decomposition product (between 35 % and 40 %) and the methane fraction increases but remains below 10 %. Mole fractions of aromatics like benzene, toluene, and styrene, increase also and in the case of cyclohexane account for nearly 30 % of decomposition products, compared to less than 5 % for n-dodecane in similar conditions.

Since jet fuels are generally a blend of linear and cyclic hydrocarbons and that the presence of cyclanes could be beneficial to the fuel properties (density, viscosity...), the effect of this higher mass fraction of aromatic compounds should be investigated because of the link between aromatic compounds and coke formation in fuel channels. Details can be found in a previous paper[6].

To compute the active cooling of such an engine, the semi-empirical NANCY code is extensively used by MBDA FRANCE. This code deals with stationary heat exchanges between two fluids flowing upstream. Implemented fluids are air, water and hydrogen, but it is possible to input other fluids (such as kerosene or endothermal hydrocarbons[7]) by defining their thermal characteristics versus their temperature and pressure. The code calculates step by step walls and fuel temperatures of each section and thus the cooling of the panels. Moreover it estimates thermal and mechanical stresses in the coolant channel material and computes the pressure drop.

By the coupling of NANCY and SENKIN codes, these computations allow to take into account the change of physical properties of the fuel with its chemical decomposition[8]. This coupling is completely operational for advanced studies (of course if the kinetic model of the fuel decomposition is available).

Another interest is the capability to check the unsteady behaviour of such a cooled structure. NANCY is a steady state code, but similar unsteady versions are under validation with same models but other 1D equations and numerical scheme[9].

But all these models are based on homogeneous pyrolysis reactions and results, while actual cooled structures, with high pressure, can not be made in quartz. Some authors have computed the heterogeneous pyrolysis[10], but no trial was done by the present team up to now.

For long basic experiments, soot formation was found after several hours of operation, and some literature surveys confirmed that elements as Fe, Ni, Cu, Ti can drastically reduce or enhance this effect[11].

The present work was then undertaken in order to obtain information about this possible effect. Focus is made here on different types of steels, commonly used on other experiments such as MPP[3]. Other basic experiment such as the COMPARER[9] one (devoted on real-time measurement techniques investigation at laboratory level) use also 316L stainless steel tubes (found to be more convenient than regular steel), but after some days of operation it was necessary to change the tubes: so they move to other materials such as titanium in order to increase the time of experiment between changing of tubes.

The present work was done to obtain some information about this possible material effect and check if the kinetic models are usable and must be updated in order to take channels material into account on the hydrocarbon chemistry.

## II. Experimental investigation of heterogeneous pyrolysis

### A. Experimental apparatus

The reactant (liquid hydrocarbon fuel) is contained in a glass vessel pressurized with nitrogen. Before performing a series of experiments, oxygen traces are removed from the liquid hydrocarbon through nitrogen bubbling and vacuum pumping. Oxygen traces could have an influence on the kinetics of the reaction.

The liquid hydrocarbon flow rate between 1 and 50 $g.hr^{-1}$ is controlled, mixed to the carrier gas (helium) and evaporated in a Controlled Mixer and Evaporator (CEM) provided by "Bronkhorst". This apparatus is composed of a liquid mass flow controller followed by a mixing chamber and a single pass heat exchanger. Carrier gas flow rate is controlled by a "3 nL.min-1 tylan – RDM280" gas mass flow controller set upstream the CEM. Molar composition of the gas at the outlet of the CEM is 2% hydrocarbon and 98% helium. Working with high dilution presents two advantages: the variation of molar flow rate due to the reaction is very weak and can be neglected and strong temperature gradients inside the reactor due to endothermic reactions are avoided.

The thermal decomposition of the fuel was performed in a quartz jet-stirred reactor that was developed by Matras and Villermaux[12]. This type of reactor which was already used for numerous gas phase kinetic studies is a spherical reactor in which diluted reactant enters through an injection cross located in its center and that can be considered to be well stirred for a residence time ($\tau$) between 0.5 and 5 s. Before entering the reactor, the gas mixture is gradually preheated to the reaction temperature in an annular preheater so that the temperature of the gas phase inside the reactor will be homogeneous.

The analysis of the products from the fuel decomposition is performed in two steps. First light species (with less than five atoms of carbon) are analyzed on-line by gas chromatography. Then heavier species are condensed in a trap dived in liquid nitrogen during an accumulation time. Then the products are analyzed by gas chromatography.

The pressure inside the jet stirred reactor is close to the atmospheric pressure (about 106 kPa). It is manually controlled thanks a control valve placed after the liquid hydrocarbon trap.

A scheme of the experimental apparatus is displayed on Figure 1.

For the present study, we used a **Catalytic Jet Stirred Reactor (CJSR)** which was developed for the investigation of hetero-homogeneous reactions[13,14]. This reactor was adapted from the spherical jet stirred reactor in order to be able to add a 40 mm diameter disk made of the material to be studied. The quartz reactor is shown in Figure 2. This reactor has a large stirred gas phase volume (~120 $cm^3$) in contact with the material disk ; the ratio surface of disk / gas reactor volume (A/V) is around 0.16 $cm^{-1}$.

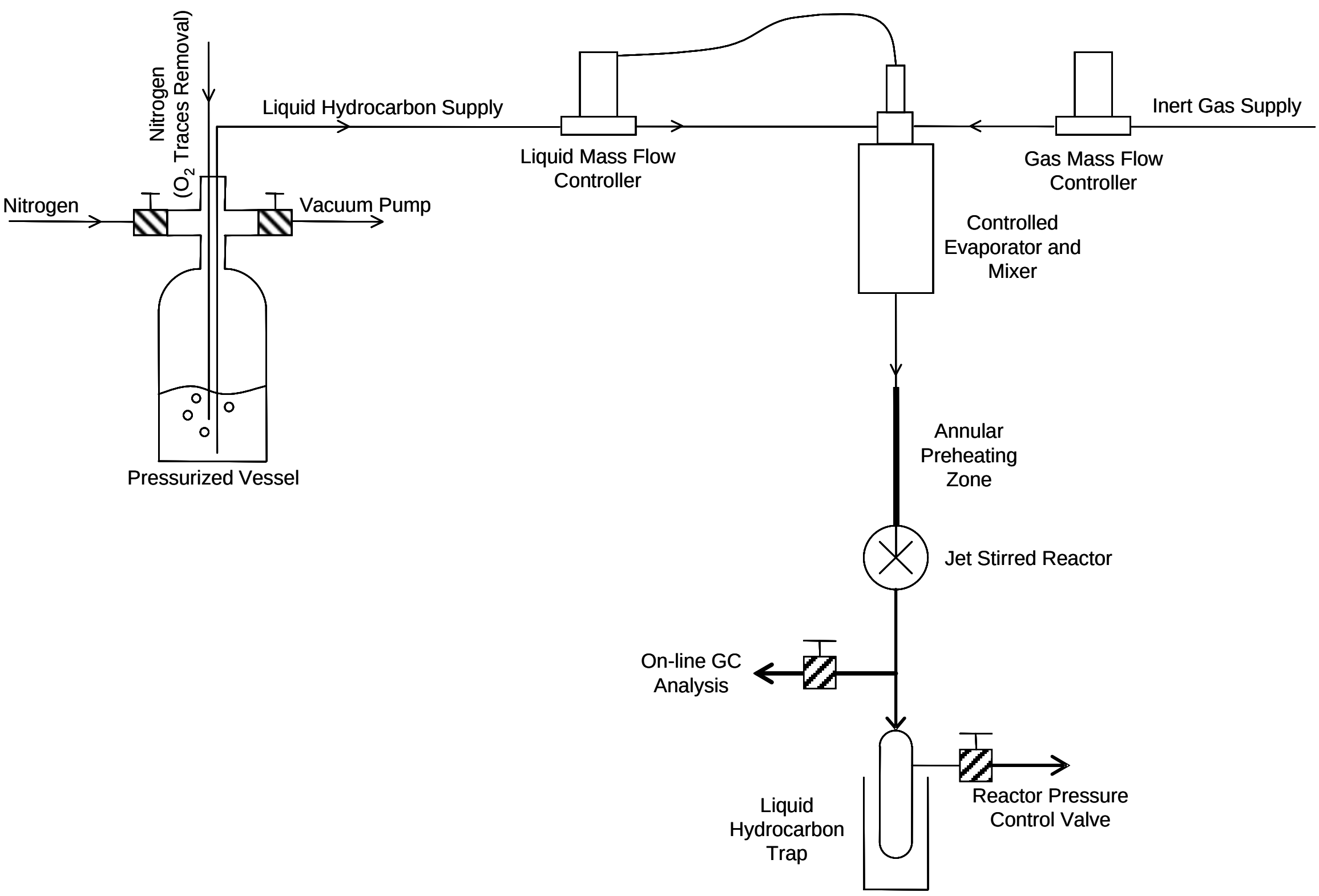

**Figure 1. Global scheme of the experimental apparatus**

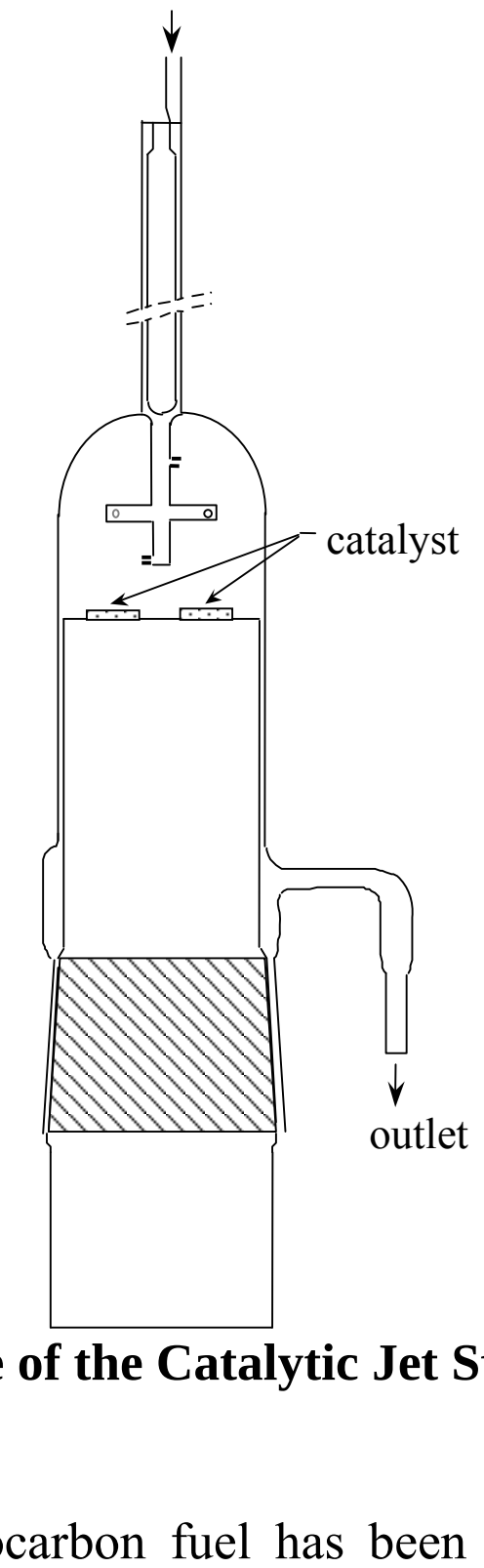

**Figure 2. Scheme of the Catalytic Jet Stirred Reactor.**

The thermal decomposition of the hydrocarbon fuel has been studied in the presence of three different materials:

- Quartz (with no effect, directly comparable to the previous homogeneous pyrolysis results).
- Regular steel (E37)
- Stainless steel (316L)

Two sets of experiments were performed with the experimental apparatus described above:

- Study 1: set of experiments at a constant residence time (3 s) for a wide range of temperatures (640 - 840°C).
- Study 2: set of experiments at a constant temperature (720°C) and residence time (3 s), but with the variation of the duration of experiments.

### B. Effect of the temperature: experimental results

In the first part of this study, we focused on the effect of the temperature on the kinetics of the reaction. Experiments have been performed in the catalytic jet stirred reactor at a constant residence time (3 s) and for a range of temperatures from 640 to 840°C. Experiments performed with the presence of a disk of material inside the reactor have been compared with homogeneous experiments (no disk inside the reactor). Three sets of experiments have been performed: two in the presence of steel and one in the presence of stainless steel.

The deposition of a solid fraction (which can be compared to soot) on the disk of material has been observed when doing experiments in the presence of steel and stainless steel. Figure 4 displays the evolution of the rate of deposition of the solid fraction on the disk with the temperature. The deposition of the solid has been quantified by weighting the disk after 30 minutes of reaction.

It can be seen that the rate of deposition is very low in the case of the stainless steel (about 10 mg/hr at 840°C) and that it is much higher in the case of steel (steel #1 in Figure 4): the rate of deposition reaches a maximum of 430 mg/hr at 740°C). In order to verify the reproducibility of the results obtained in the presence of steel, a second set of experiments was performed (steel #2 in Figure 4). The rate of deposition of the solid fraction is still higher than the one for stainless steel but it is different from the one for steel #1 (these experiments exhibit a maximum disposition rate of 160 mg/hr at 720°C). An explanation could be a slight difference in the composition of the two sets of disks or a difference in the texture of the surface of the disks (roughness). Further investigations (e.g. analysis of the elementary composition, analysis of the roughness of the surface) are required to find an answer to this question.

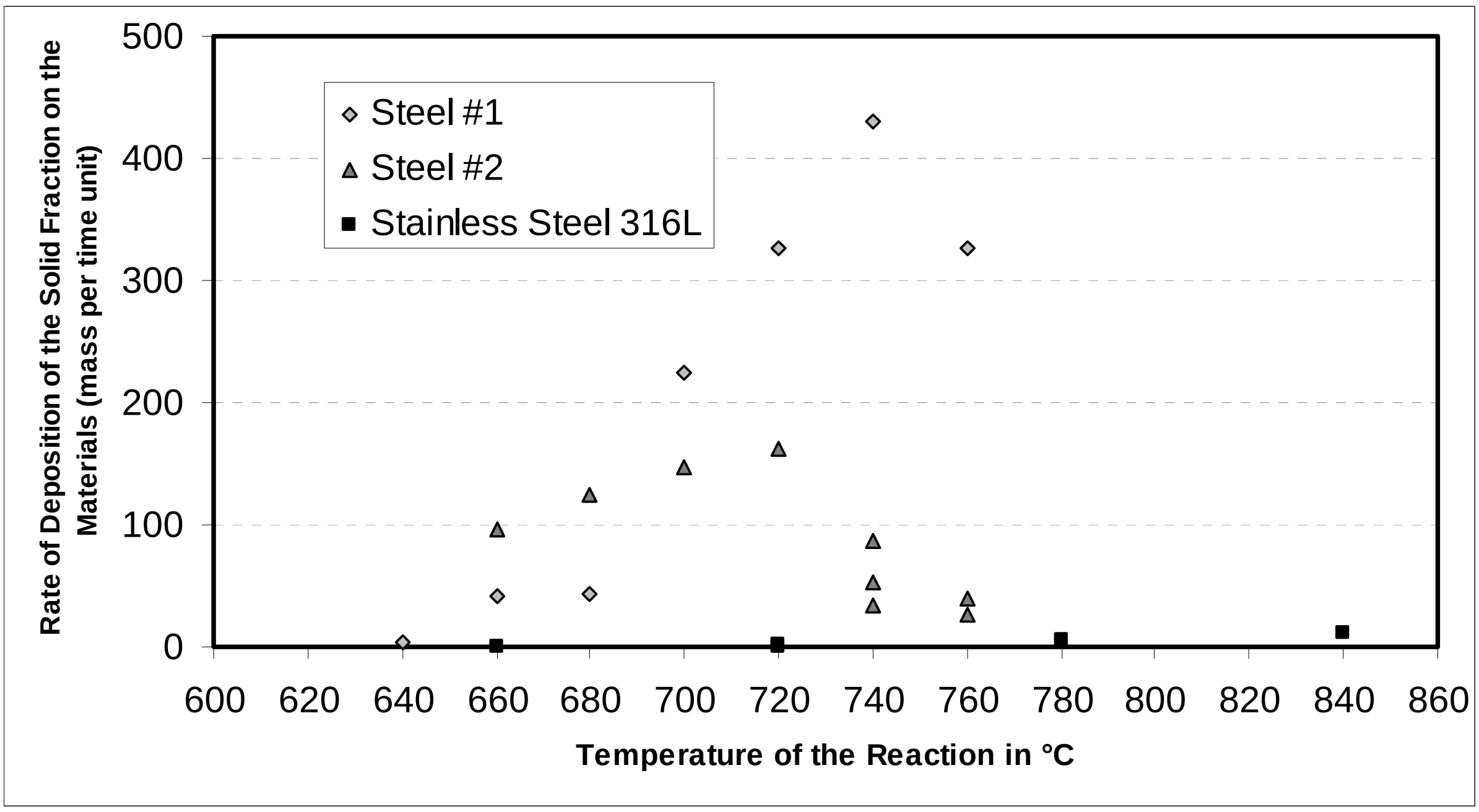

**Figure 4. Rate of deposition of the solid fraction on the disks ($\tau$ = 3 s).**

Comparisons for the mole fractions of the reactant and of some products of the reaction in the gas phase (methane, hydrogen) are displayed on Figures 5, 6 and 7 for experiments performed in the presence of steel #2 and in the absence of any material (homogeneous pyrolysis).

As far as the conversion of the reactant is concerned, it can be seen on Figure 5 that there are very little differences between the two sets of experiments. At the highest temperatures (above 720°C), conversions obtained in the presence of steel are a little bit less than conversions in homogenous pyrolysis. This may be due to an inhibiting effect but given the slight difference between the values further experiments are required for confirmation.

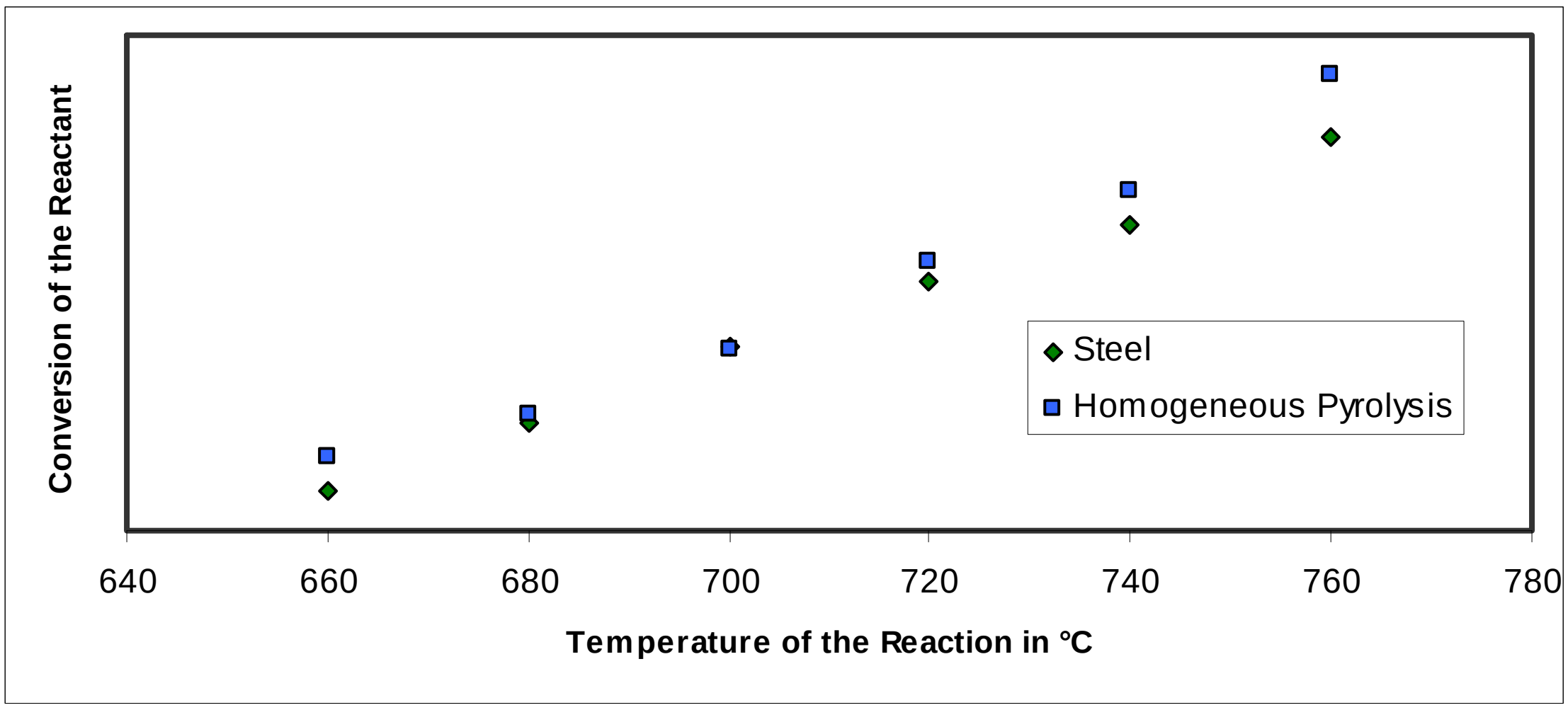

**Figure 5. Conversion of the reactant with and without steel inside the reactor.**

The mole fractions of hydrogen and methane in the gas phase are very little affected by the presence of steel (Figures 6 and 7). It can be seen of Figure 6 that hydrogen mole fractions obtained in the presence of steel are higher at 700 and 720°C. In the case of methane there is no difference between the two sets of experiments except at the temperature of 740°C (that was also the case for hydrogen; the results obtained at this temperature are likely wrong).

There are also very little differences in the mole fractions of the other products of the reaction in the gas phase obtained in the presence of the steel and in homogeneous pyrolysis.

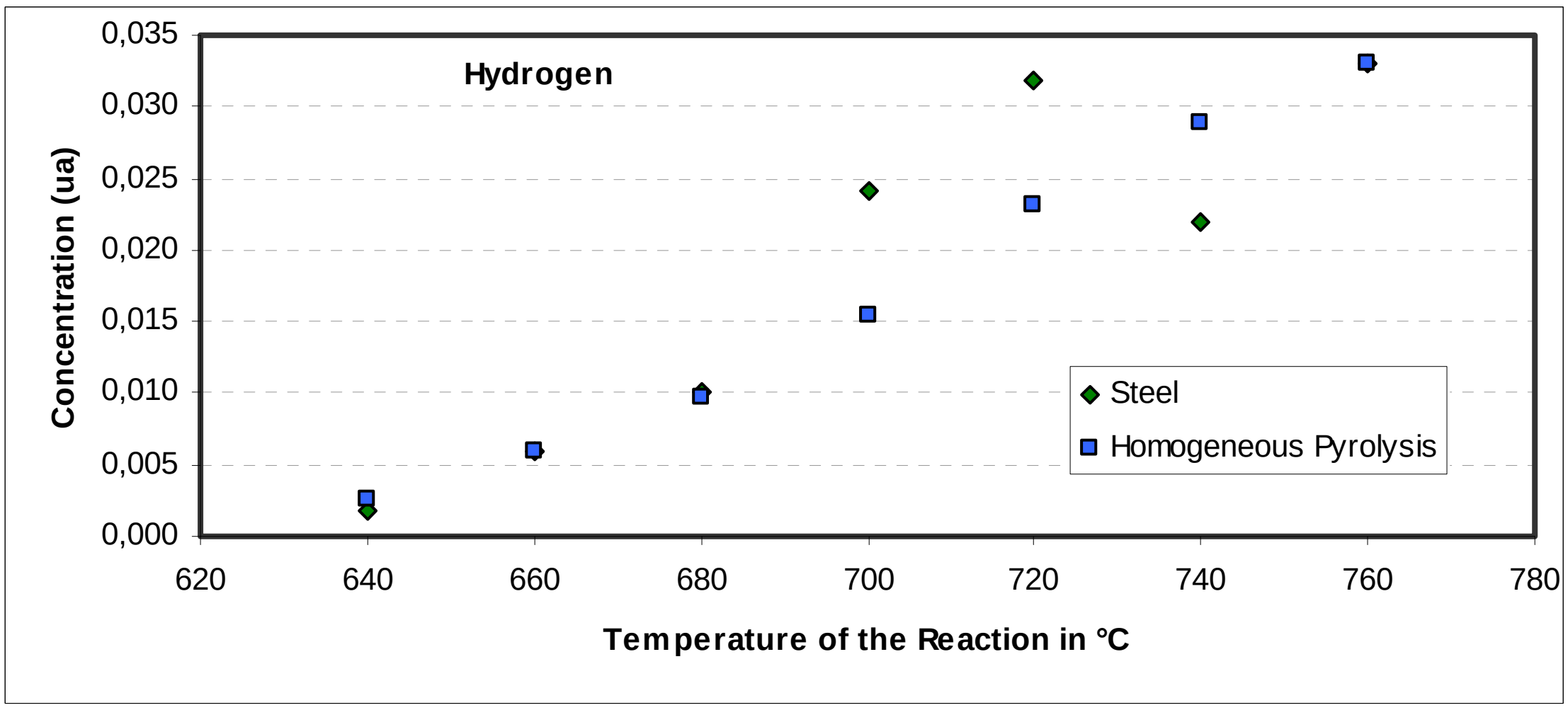

**Figure 6. Evolution of the mole fraction of hydrogen with the temperature for experiments performed with and without steel inside the reactor.**

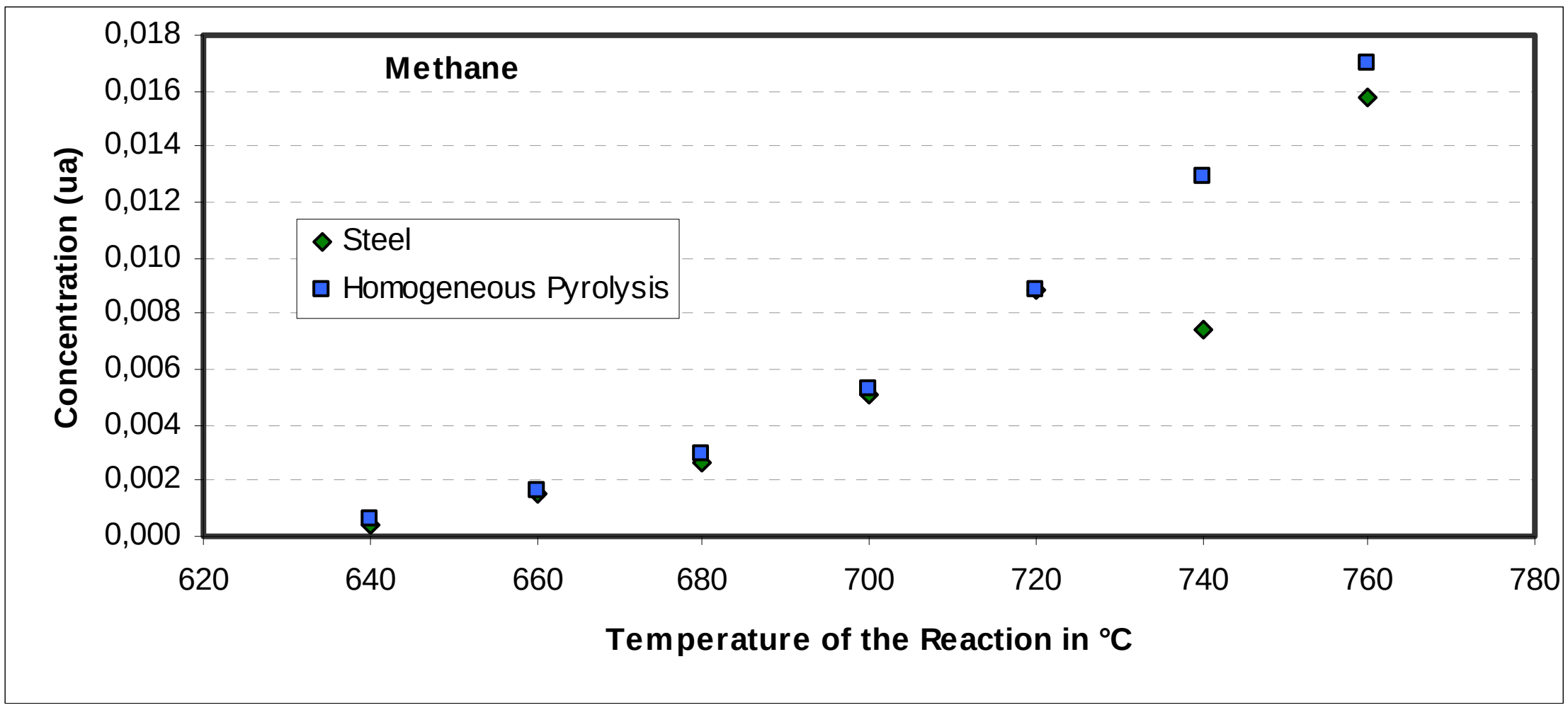

**Figure 7. Evolution of the mole fraction of methane with the temperature for experiments performed with and without steel inside the reactor.**

During the first part of this first study, the effect of different materials (steel and stainless steel) on the kinetics of the reaction of decomposition of the fuel has been investigated. The deposition of a solid on the disks has been observed with both materials but the rate of deposition is much higher with steel than with stainless steel.

In the case of experiments conducted with steel, we obtained a maximum rate of deposition at a temperature around 720 – 740°C. The results obtained with two different sets of disks of steel are different although the disks are assumed to be made from the same materials (steel E37). Further investigations are necessary to explain this fact.

## C. Effect of the duration of experiments: results

In the first part of the study, the effect of different materials on the kinetic of the reaction has been studied for a wide range of temperatures (640 – 840°C) and at a constant residence time (3 s). The deposition of a solid fraction on the disks has been observed.

But this effect due to the presence of the disk in the reactor is likely evolving with the duration of the experiments due the deposition of the solid fraction at the surface of the disk. So in the second part of the study new experiments have been performed at a constant temperature of 720°C (corresponding to a maximum rate of deposition of solid on the disk with steel #2), at a constant residence time (3 s) and by varying the duration of the experiment.

The analysis of the products of the reaction in the gas phase and the weighing of the solid deposit on the disk have been performed every 5 minutes. Two sets of identical experiments have been performed because it is not possible to carry out the online analysis of the species at the same time than the accumulation of the heaviest species in the trap dived in the liquid nitrogen.

Figure 8 displays the evolution of the conversion of the reactant with the duration of the experiments in the presence of steel #2 and stainless steel and in the absence of any material. It can be seen on this figure that the duration of the experiments has very little effect on the conversion of the reactant whatever the material.

In the conditions of the experiments (T = 720°C and $\tau$ = 3 s), the deposition of a solid fraction on the disk has been observed in the case of steel #2 only. No deposit was observed with stainless steel in accordance with the results obtained in the first part of the study (the rate of deposition at 720°C is very low in the presence of stainless steel). Figure 9 displays the evolution of the mass of deposit on the disk of steel #2 with the duration of the experiment. It can be seen on this graph that the mass of deposit increases rapidly in the early time of the experiment. The increase in the mass becomes lower when the duration of the experiment is important. These results show that the deposition of the solid at the surface of the disk has an effect on the kinetics of deposition (the nature and the texture of the surface is changing due to the deposition of the solid). The evolution of the rate of deposition of the solid on the disk is shown in Figure 10. In accordance with the results shown in Figure 9, the rate of deposition is very important in the early time of the reaction and goes down rapidly when the duration of the experiment increases.

Figure 11 displays the evolution of the mole fraction of hydrogen with the duration of the experiments. It is interesting to notice that there is no difference between the homogeneous experiments and the experiments performed in the presence of stainless steel, whereas the mole fraction of hydrogen obtained in the presence of steel #2 is much higher for experiment duration less than 20 minutes. For higher experiment durations the mole fraction of hydrogen is very close to the ones obtained during the two other sets of experiments.

It seems that these higher hydrogen mole fractions at the lowest experiment duration are correlated to the rate of deposition of the solid on the disk of steel #2. An explanation could be that the formation of hydrogen is linked to the deposition of the solid deposit on the disk. This solid deposit is likely soot which has a higher carbon to hydrogen atoms ratio than the reactant and most products of the reaction.

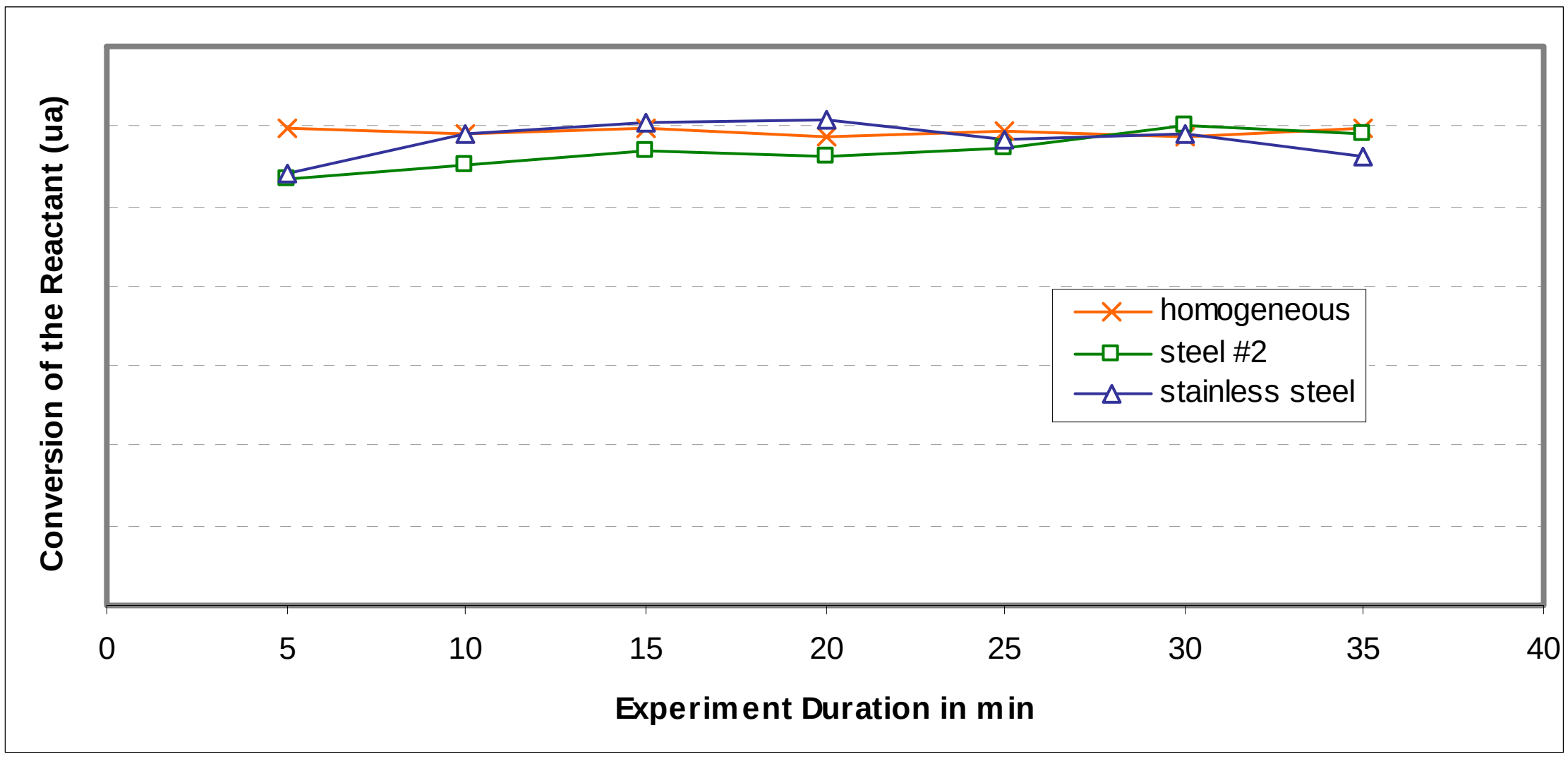

**Figure 8. Conversion of the reactant with the duration of the experiments (T = 720°C, τ = 3s).**

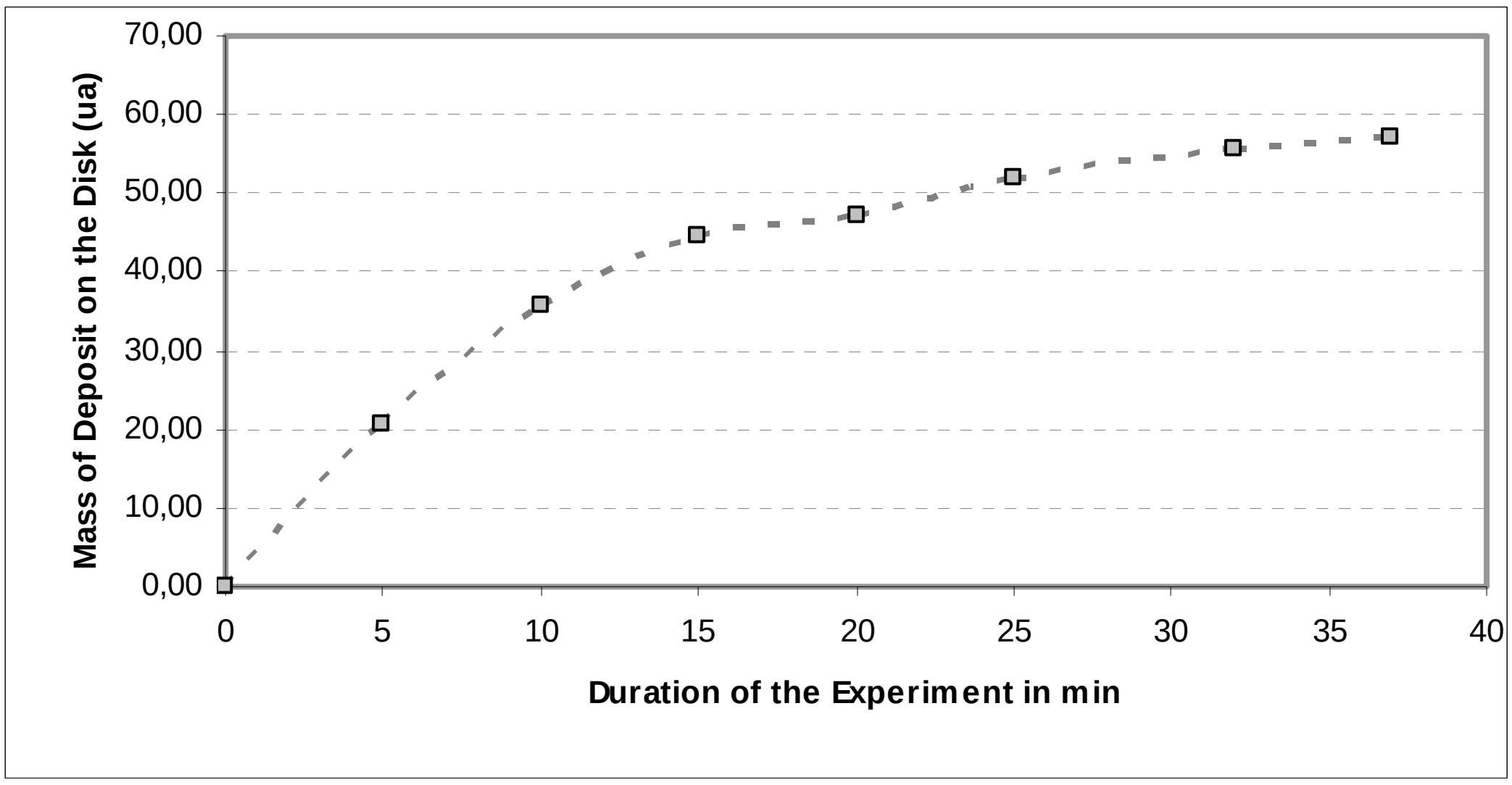

**Figure 9. Evolution of the mass of the solid deposit on the disk of steel with the duration of the experiments (steel #2, T = 720°C, τ = 3s).**

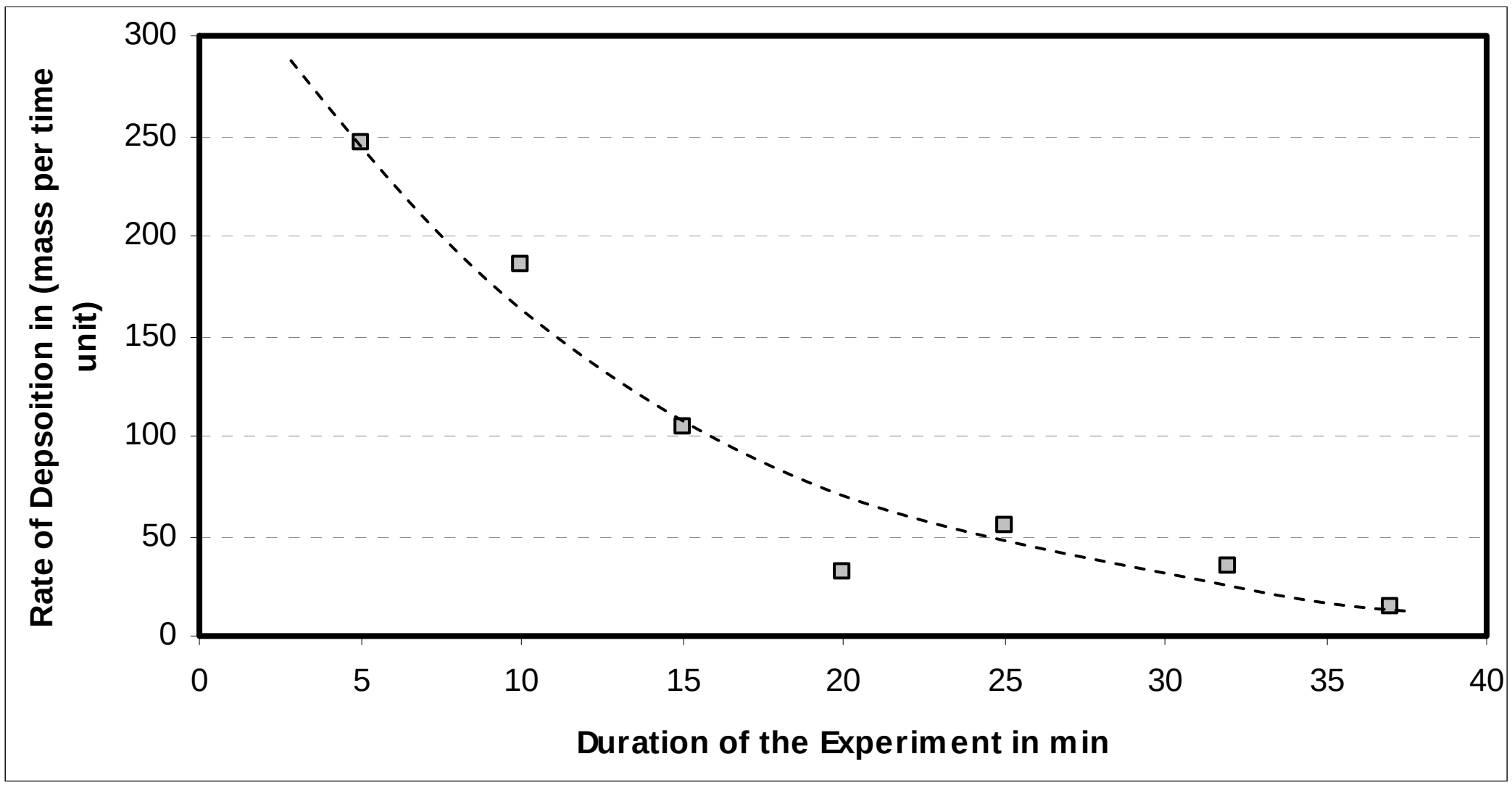

**Figure 10. Evolution of the rate of deposition of the solid fraction on the disk.**

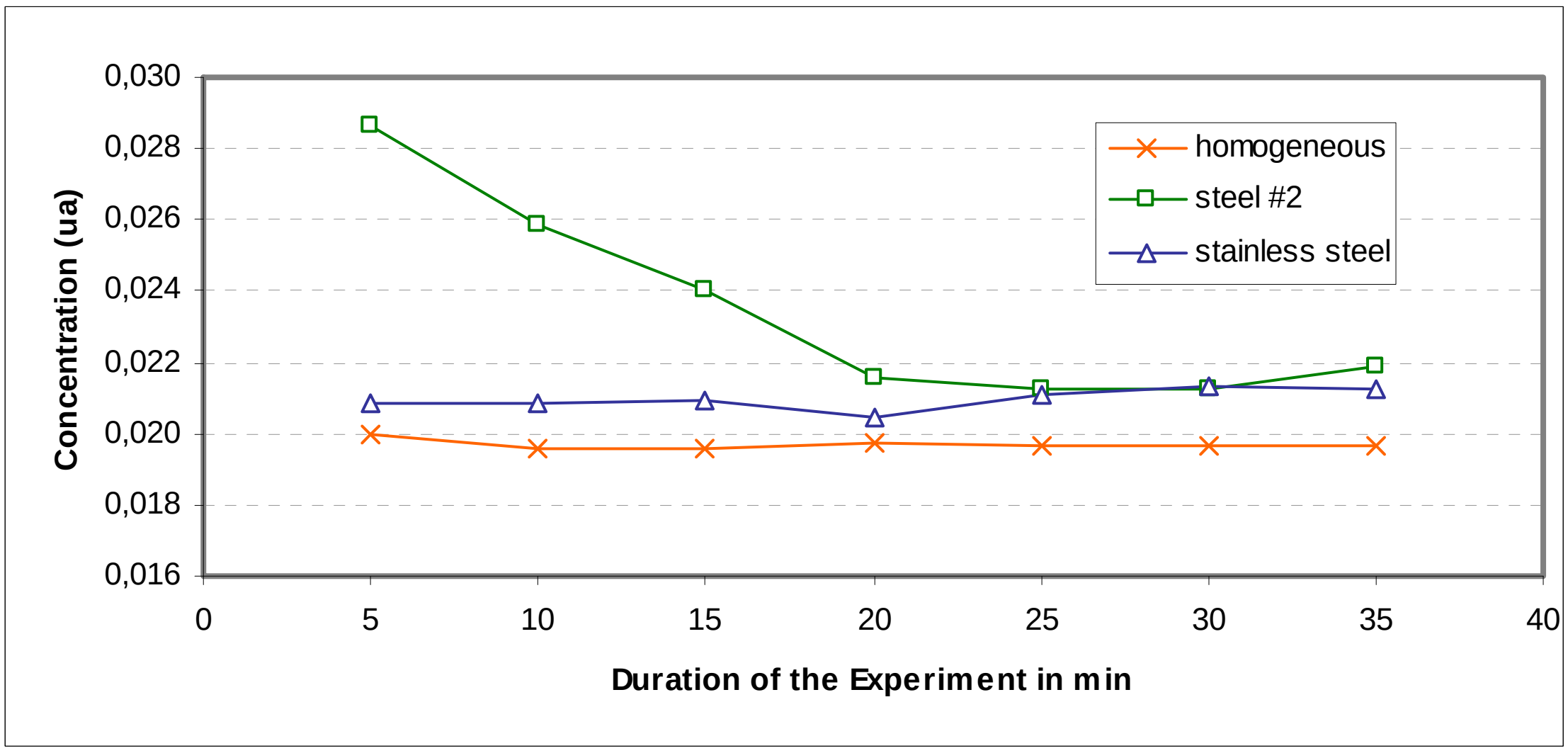

**Figure 11. Evolution of the mole fraction of hydrogen with the duration of the experiment.**

The results obtained in the second part of the study (evolution with the contact time between the gas phase and the disk inside the reactor) lead to very interesting observations. The conversion of the reactant and the mole fractions of most of the products of the reaction are not affected by the duration of the experiment. The main differences are visible for the deposition of the solid fraction on the disk of steel #2 and for the mole fraction of hydrogen. The rate of deposition of the solid on the disk evolves a lot with the duration of the experiment and it seems that the evolutions of both the rate of deposition and the mole fraction of hydrogen are correlated.

## III. Conclusion

The effect of steel or stainless steel on the thermal decomposition of a hydrocarbon fuel has been studied in a catalytic jet-stirred reactor over a wide range of temperatures from 640 to 840°C and for a residence time of 3 s.

As far as the conversion of the reactant is concerned, no important effect has been observed when experiments were conducted in the presence of a disk of steel or stainless steel inside the reactor. Thus homogeneous pyrolysis kinetic model can still be used to do simulations in actively cooled structures.

We could observe that the presence of stainless steel has very little influence on the rate of deposition of the solid whereas it is very important with steel. In the case of this last material, the study shows that the rate of deposition of the solid on the disk goes down rapidly when the duration of the experiment increases. In the conditions of the study, the effect of the presence of the disk in the reactor is not visible anymore from 20 minutes. This let us think that the deposit has no catalytic effect on the reaction of decomposition of the fuel and that the material has no influence anymore once it is covered by the solid deposit.

These new results must be checked in more actual conditions and can only be considered like a preliminary study to pave the way of understanding the use of hydrocarbon on regeneratively cooled engines.

---

[1] Dufour, E., and Bouchez, M., *AIAA (2002), 5126.*

[2] Daniau, E., Bouchez, M., Herbinet, O., Marquaire, P.-M., Gascoin, N., and Gillard, P., *AIAA (2005), 3403.*

[3] Sickard, M., Raepsaet, B., Ser, F., and Masson, C., *AIAA (2006) 7974.*

[4] O. Herbinet, PM. Marquaire, F. Battin-Leclerc, R. Fournet, "Thermal stability of n-dodecane : experiments and kinetic modeling", J. Anal. Appl. Pyrol., vol 78, Issue 2, p 419-429 (2007)

[5] O. Herbinet, B. Sirjean, F. Battin-Leclerc, R. Fournet and PM. Marquaire, "Thermal decomposition of norbornane (bicyclo[2.2.1]heptane). Experimental study and mechanism investigation", Energy & Fuels vol 21,n°3 p 1406-1414 (2007)

[6] E. Daniau et al. « Numerical simulations and experimental results of endothermic fuel reforming for scramjet cooling application", AIAA-2006-7945, Canberra, nov 2006.

[7] F. Ser, B. Heinrich, A. Luc-Bouhali, "Carburants liquides endothermiques : problématique du refroidissement et projet d'études expérimentales à l'ONERA », AAAF – Arcachon, France – March 2001

[8] E. Daniau, M. Bouchez, R. Bounaceur, F. Battin-Leclerc, P.-M. Marquaire, R. Fournet, "Contribution to Scramjet active cooling analysis using n-dodecane decomposition model as a generic endothermic fuel". AIAA-2003-6920, December 2003, Norfolk, USA.

[9] N. Gascoin, P. Gillard, S. Bernard, Y. Touré, E. Daniau, E. Dufour, M. Bouchez, "Transient Numerical Model of Scramjet Active Cooling, Application to an Experimental Bench", 4th International Energy Conversion Engineering Conference and Exhibit (IECEC), 26-29 June 2006, San Diego, USA

[10] A.S. Wade et al., « Optimisation of the Arrhenius parameters in a semi-detailed mechanism for jet fuel thermal degradation using a genetic algorithm », GT2004-53367, presented at ASME turbo Expo 2004, Vienna, Austria.

[11] J. Weill, P. Broutin, F. Billaud, C. Gueret, "Coke formation during hydrocarbons pyrolysis_part one: steam cracking", Oil & Gas Science Technology, Revue de l'Institut Français du Pétrole, 1992, vol 47 (4), 537-549.

[12] Matras, D., and Villermaux*, J., Chem. Eng. Sci., 28 (1973), 129.*

[13] I. Ziegler, R. Fournet, P. M. Marquaire, "Influence of surface on chemical kinetic of pyrocarbon deposition obtained by propane pyrolysis", J. Anal. Appl. Pyrol., vol 73-1, pp 107-115 (2005)

[14] M. Fleys, W. Shan,Y Simon, P.M. Marquaire, "A detailed kinetic study of the reaction of partial oxidation of methane over La2O3 catalyst.-Part 1 : Experimental results", Ind. Eng. Chem. Res., vol 46, issue 4, pp 1063-1068 (2007).